\documentclass[manuscript]{aastex}

\usepackage{graphicx}
\usepackage{epstopdf}
 \usepackage{natbib}

\slugcomment{Draft - version 20110223\_Proof}

\shorttitle{VHE DETECTION FROM THE FSRQ PKS\,1222+21}
\shortauthors{MAGIC coll.}

\begin{document}

\title{MAGIC discovery of Very High Energy Emission from the FSRQ PKS\,1222+21}
%
\author{
J.~Aleksi\'c$^{1}$,
L.~A.~Antonelli$^{2}$,
P.~Antoranz$^{3}$,
M.~Backes$^{4}$,
J.~A.~Barrio$^{5}$,
D.~Bastieri$^{6}$,
J.~Becerra Gonz\'alez$^{7,8,\dag}$,
W.~Bednarek$^{9}$,
A.~Berdyugin$^{10}$,
K.~Berger$^{7,8}$
% ...
\\
(MAGIC Collaboration)\\
Y.~T.~Tanaka$^{30}$,
D.~L.~Wood$^{31}$,
S.~Buson$^{6}$
}
\altaffiltext{$^{1}$} {IFAE, Edifici Cn., Campus UAB, E-08193 Bellaterra, Spain}
\altaffiltext{$^{2}$} {INAF National Institute for Astrophysics, I-00136 Rome, Italy}
\altaffiltext{$^{3}$} {Dipartimento di Fisica, Universit\`a  di Siena, and INFN Pisa, I-53100 Siena, Italy}
\altaffiltext{$^{4}$} { Fakult\"at f\"ur Physik, Technische Universit\"at Dortmund, D-44221 Dortmund, Germany}
\altaffiltext{$^{5}$} {Grupo de Fisica Altas Energias, Universidad Complutense, E-28040 Madrid, Spain}
\altaffiltext{$^{6}$} {Dipartimento di Fisica, Universit\`a di Padova and INFN, I-35131 Padova, Italy}
\altaffiltext{$^{7}$} {Inst. de Astrof\'{\i}sica de Canarias, E-38200 La Laguna, Tenerife, Spain}
\altaffiltext{$^{8}$} {Depto. de Astrof\'{\i}sica, Universidad de La Laguna, E-38206 La Laguna, Spain}
\altaffiltext{$^{9}$} {Division of Experimental Physics, University of \L\'od\'z, PL-90236 Lodz, Poland}
\altaffiltext{$^{10}$} {Tuorla Observatory, University of Turku, FI-21500 Piikki\"o, Finland}
% ...
\altaffiltext{$^{30}$} {Institute of Space and Astronautical Science, JAXA, 3-1-1 Yoshinodai, Chuo-ku, Sagamihara, Kanagawa 252-5210, Japan}
\altaffiltext{$^{31}$} {Space Science Division, Naval Research Laboratory, Washington, DC 20375, USA}
\altaffiltext{\dag} {Send offprint requests to J.~Becerra Gonz\'alez (jbecerra@iac.es),  D.~Mazin (mazin@ifae.es), K.~Saito (ksaito@mpp.mpg.de) and A.~Stamerra (antonio.stamerra@pi.infn.it)}


\date{Received 2011 January 22;  Accepted 2011 February 11; published 2011 March 20}
\begin{abstract}
Very high energy (VHE) $\gamma$-ray emission from the flat spectrum radio quasar (FSRQ) PKS\,1222+21 (4C\,21.35, $z$=0.432) was detected with the MAGIC Cherenkov telescopes during a short observation ($\sim$0.5\,hr) performed on 2010 June 17. 
 The MAGIC detection coincides with high energy MeV/GeV $\gamma$-ray activity measured by the Large Area Telescope (LAT) on board the \textit{Fermi} satellite.
 The VHE spectrum measured by MAGIC extends from about 70\,GeV up to at least 400\,GeV 
 and can be well described by a power law $d{\rm N}/d{\rm E}\propto{\rm E}^{- \Gamma }$  with a photon index  $\Gamma=3.75\pm0.27_{\tiny \mbox{stat}}\pm0.2_{\tiny \mbox{syst}}$. 
 The averaged integral flux above 100\,GeV is $(4.6\pm0.5) \times 10^{-10}\,\mbox{cm}^{-2}\mbox{s}^{-1}$
 ($\sim1$ Crab Nebula flux).  
 %
 The VHE flux measured by MAGIC varies significantly within the 30\,minutes exposure implying a flux doubling time of about 10\,minutes.
The VHE and MeV/GeV spectra, corrected for the absorption by the extragalactic background light (EBL), can be described by a single power law with photon index $2.72\pm0.34$ between 3\,GeV and 400\,GeV, and is consistent with emission belonging to a single component in the jet.
   The absence of a spectral cutoff constrains the  $\gamma$-ray emission region to lie outside the broad-line region, which would otherwise absorb the VHE $\gamma$-rays.
  Together with the detected fast variability, this challenges present emission models from jets in FSRQs.
  Moreover, the combined \textit{Fermi}/LAT and MAGIC spectral data yield constraints on the density of the EBL in the UV-optical to near-infrared range that are compatible with recent models.
\end{abstract}

\section{Introduction}
High-luminosity active galactic nuclei (AGNs) hosting  powerful relativistic jets are characterized by strong nonthermal emission extending across the entire electromagnetic spectrum, from radio up to $\gamma$-rays.
More than 40 AGNs have been detected in the very high energy (VHE) domain ($E>$\,100\,GeV) by ground based 
Cherenkov telescopes\footnote{For an updated list refer to http://tevcat.uchicago.edu/ or {http://www.mppmu.mpg.de/$\sim$rwagner/sources/} }. The great majority of them are BL Lac objects, while only two are classified as flat spectrum radio quasars (FSRQs): PKS 1510--08 \citep[$z=0.36$,][]{1510HESS} and 3C\,279 \citep[$z=0.536$,][]{MAGIC3c279,3C279magic2011}, the most distant VHE source detected up to now. 

FSRQs display luminous, broad emission lines often accompanied by a ``big blue bump" in the optical-UV region, associated with the direct emission from the accretion disk. 
VHE emission from FSRQs may therefore be affected by internal absorption from the dense UV-optical radiation
reprocessed in the Broad Line Region (BLR)~\citep{Donea2003}.
Distant  VHE quasars offer the possibility to probe the Extragalactic Background Light (EBL), the integrated stellar and dust emission through cosmic history, in the range 0.1--10 $\mu$m  \citep{HauserDwek2001}.

The MAGIC detection \citep{MAGIC-ATel} of the FSRQ PKS\,1222+21 \citep[4C 21.35, $z=0.432$,][]{Osterbrock1987}  makes it the second most distant object with known redshift (after 3C 279) detected at VHE\footnote{\small {The redshift measurement ($z=0.444$) of the VHE BL Lac 3C 66A has large uncertainties~\citep{Bramel2005}.}}. 
PKS\,1222+21 is a $\gamma$-ray blazar \citep{Abdo1LBAS} with a relatively hard spectrum in the GeV range and has been included in the list of $>$100~GeV emitters in the analysis of \citet{Neronov}. It is characterized by highly superluminal jet knots with apparent velocity up to 21\,{\it c} \citep{Lister2009}.

Upper limits on the VHE emission of PKS\,1222+21 have been previously derived by Whipple \citep{ULWhipple} at the level of $12\times10^{-12}\mbox{cm}^{-2}\,\mbox{s}^{-1}$ at E$>300$\,GeV.
We report here on the MAGIC discovery of this source,  during a phase of high activity in $\gamma$-rays announced by the \textit{Fermi}/LAT collaboration. We discuss its implications for the EBL studies and the blazar physics.\\

\section{Observations }

MAGIC consists of two 17\,m diameter Imaging Atmospheric Cherenkov Telescopes
(IACT) located at the Roque de los Muchachos, Canary Island of La Palma ($28^{\circ}46'\,$N,
$17^{\circ}53'\,$W), at the height of 2200\,m a.s.l. 
The stereo observations provide a sensitivity
\footnote{\small{Sensitivity is defined here as the minimal integral flux to reach 5$\sigma$ signal in 50\,h of observations.}}
 of $0.8\%$  of the Crab nebula flux at E$>250\,$GeV~\citep{colinStereo}.

PKS\,1222+21 was observed by MAGIC from May 3 to June 19 2010 (MJD 55319 to MJD 55366) for a total of $\sim$14.3\,h. 
The observations started as a part of a Target of Opportunity program triggered by an increase of the flux in the \textit{Fermi} passband~\citep{FermiATel}. 
In this letter we report the results obtained from the observation of
June 17 (MJD 55364), when the source was detected by MAGIC in close coincidence with 
the brightest flare observed by \textit{Fermi} Large Area Telescope (LAT)~\citep{Fermi draft}.
Results from the multi-wavelength campaign covering all 2010 observations will be published elsewhere. 
Nevertheless a preliminary analysis does not provide any high-significant detection with MAGIC in any other day during the campaign.

On June 17, 21:50 UT, PKS\,1222+21 was observed with the MAGIC telescopes for $\sim$0.5\,h (MJD\,55364.908 to MJD\,55364.931), in the so-called wobble mode.
The data were taken at zenith angles between $26^\circ$ and $35^\circ$.
The light conditions during the observations correspond to moderate moon light leading to a higher noise level in the data. A cleaning level higher than the standard one was therefore applied to remove signals from night sky background noise.
Stereoscopic events, triggered by both MAGIC telescopes, were analyzed in the MARS analysis framework~\citep{MoralejoLodz}. 
Details on the analysis can be found in \citet{3C66Amagic2010} whereas the performance of the MAGIC telescope stereo system will be discussed in detail in a forthcoming paper.


\section{Results}

The strength of the signal was evaluated applying standard cuts to the PKS\,1222+21 data sample, corresponding to an energy threshold of $\approx 70\,$GeV as determined by Monte Carlo events, assuming a soft spectrum with a photon index of $\Gamma = 3.5$. The $\theta^2$ distribution (squared angular distance between the true and reconstructed source position) of the signal coming from the region of PKS\,1222+21 yields an excess of 190 $\gamma$-like events (6\,$\gamma$/min.),  corresponding to a statistical significance of 10.2\,$\sigma$ using eq.\ 17 in \citet{LiMa}. 

\begin{figure}[tbp]
\plotone{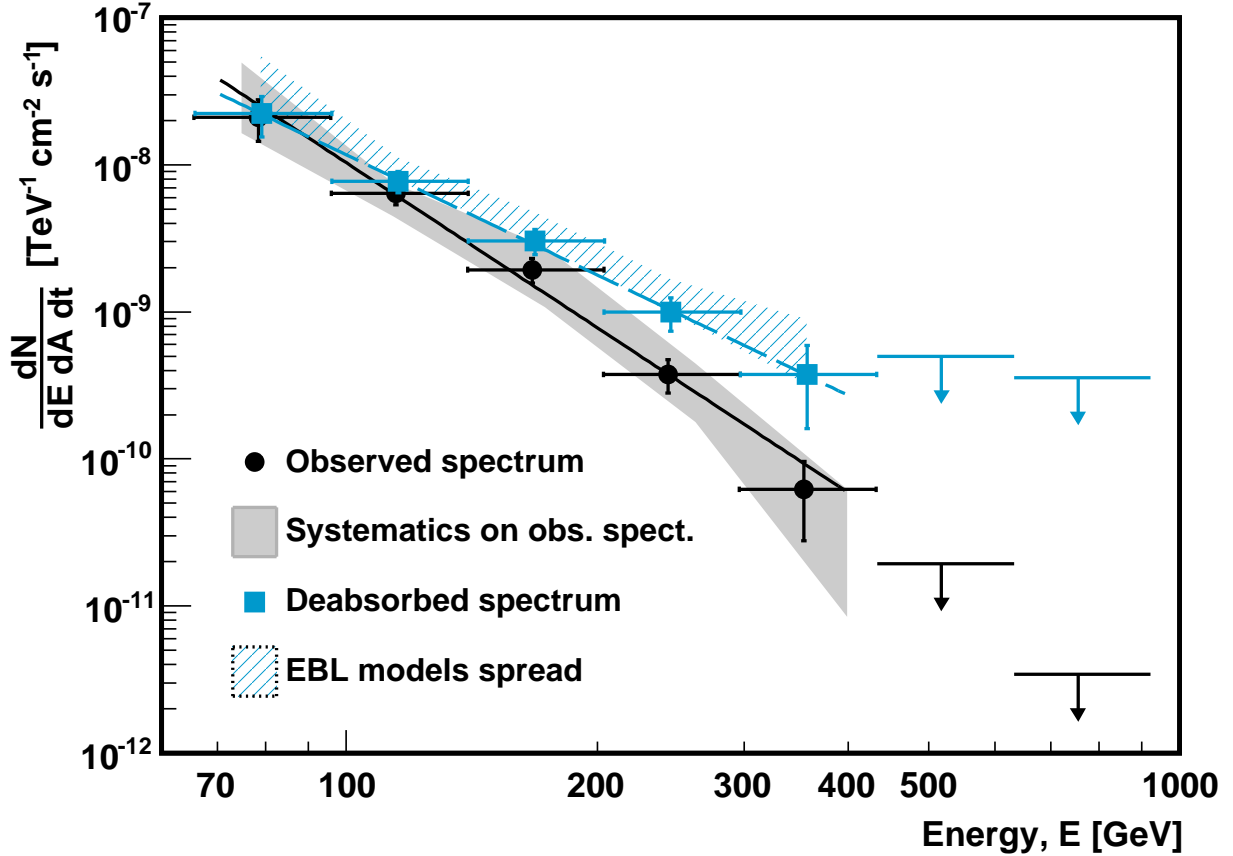}
\caption{Differential energy spectrum of PKS\,1222+21 as 
measured by MAGIC on 2010 June 17. Differential fluxes are shown as black points, 
upper limits (95\% C.L.)  as  black arrows. The black line is the best fit to a power law.
The grey shaded area represents the systematic uncertainties of the analysis. 
The absorption corrected spectrum and upper limits using the EBL model by
\citet{Dominguez2010} are shown by the blue squares and arrows; the dashed blue line is the best fit power law.
The blue-striped area illustrates the uncertainties  due to differences in the EBL models cited in the text.
by \citet{kneiskedole2010,gilmore:2009a,franceschini} and \citet{MAGIC3c279}.
}
\label{fig:spectrum}
\end{figure}

\subsection{VHE spectrum}

The differential energy spectrum of PKS\,1222+21 was reconstructed using the ``Tikhonov" unfolding algorithm \citep{unfolding}, to take into account the finite energy resolution of the instrument and the biases in the energy reconstruction.
The energy spectrum,  shown in Fig.~\ref{fig:spectrum},  extends up to at least 400\,GeV and is well-described by a simple power law of the form:
\begin{equation}
\frac{dN}{dE}=N_{200}\left(\frac{E}{200\,\mbox{GeV}}\right)^{-\Gamma}
\end{equation}
with a photon index $\Gamma=3.75 \pm 0.27_{\tiny\mbox{stat}}\pm0.2_{\tiny\mbox{syst}}$ and a normalization constant at 200\,GeV of 
$N_{\tiny 200}=(7.8\pm 1.2_{\tiny\mbox{stat}}\pm3.5_{\tiny\mbox{syst}})\times10^{-10}\mbox{cm}^{-2}\mbox{s}^{-1}  \mbox{TeV}^{-1}$, yielding an integral flux 
$(4.6\pm0.5)\times 10^{-10}\,\mbox{cm}^{-2}\mbox{s}^{-1}$ ($\approx 1$ Crab Nebula flux) at E$>100\,$GeV and $(9.0\pm3.6)\times 10^{-12}\,\mbox{cm}^{-2}\mbox{s}^{-1}$ (7\% of the Crab Nebula flux) at E$>300$\,GeV,  
at the same level of Whipple upper limit~(sec.1). 
For energies higher than 400\,GeV no significant excess was measured. The upper limits corresponding to 95\% confidence level (C.L.) are shown in Fig.~\ref{fig:spectrum}.
The systematic uncertainty of the analysis (studied by using different cuts and different unfolding algorithms) is shown by the grey area. 

We studied the effect of the VHE $\gamma$-ray absorption due to pair-production with 
low energy photons of the EBL by using different state-of-the-art EBL models, namely the models by 
\citet{Dominguez2010, kneiskedole2010,gilmore:2009a, franceschini} and the ``max high UV" EBL model described in \citet{MAGIC3c279}.
For each of the EBL models the  optical depth corresponding to  the measured VHE $\gamma$-ray energy intervals was computed and the differential fluxes were corrected accordingly to obtain the de-absorbed (or intrinsic) spectrum. 
 The  spectrum deabsorbed with the EBL model of  \cite{Dominguez2010}, 
 shown by the blue squares in Fig.~\ref{fig:spectrum},  are well fitted  by a power law with an intrinsic photon index of $\Gamma_{\tiny\mbox{intr}}=2.72\pm 0.34$
 between 70\,GeV and 400\,GeV. 
Uncertainties caused by the differences between the EBL models are represented in Fig.~\ref{fig:spectrum} by the blue-striped area. The corresponding spread is smaller than the systematic uncertainties of the MAGIC data analysis.

We investigated the possible presence of a high energy cut off in the VHE range
by fitting power laws with different photon indexes and different
values for the cut off. The method 
adopted is the $\chi^2$ difference method \cite[see, e.g.][]{Lampton1976}.
With the available statistics, at the 95\% C.L. we cannot exclude 
the presence of a cut off above 130 GeV for a photon index 2.4 (the lowest possible value compatible 
with fit uncertainties and with \textit{Fermi}/LAT data, see sec. 3.3) 
or above 180 GeV for a photon index 2.7. The confidence interval
is not bounded on the high energy side, i.e. a fit without a cutoff
is fully compatible with the data.
Further observations with higher statistics are needed to better constrain
the location of a possible steepening in the form of a cut off or spectral break.

\begin{figure}[bp]
\plotone{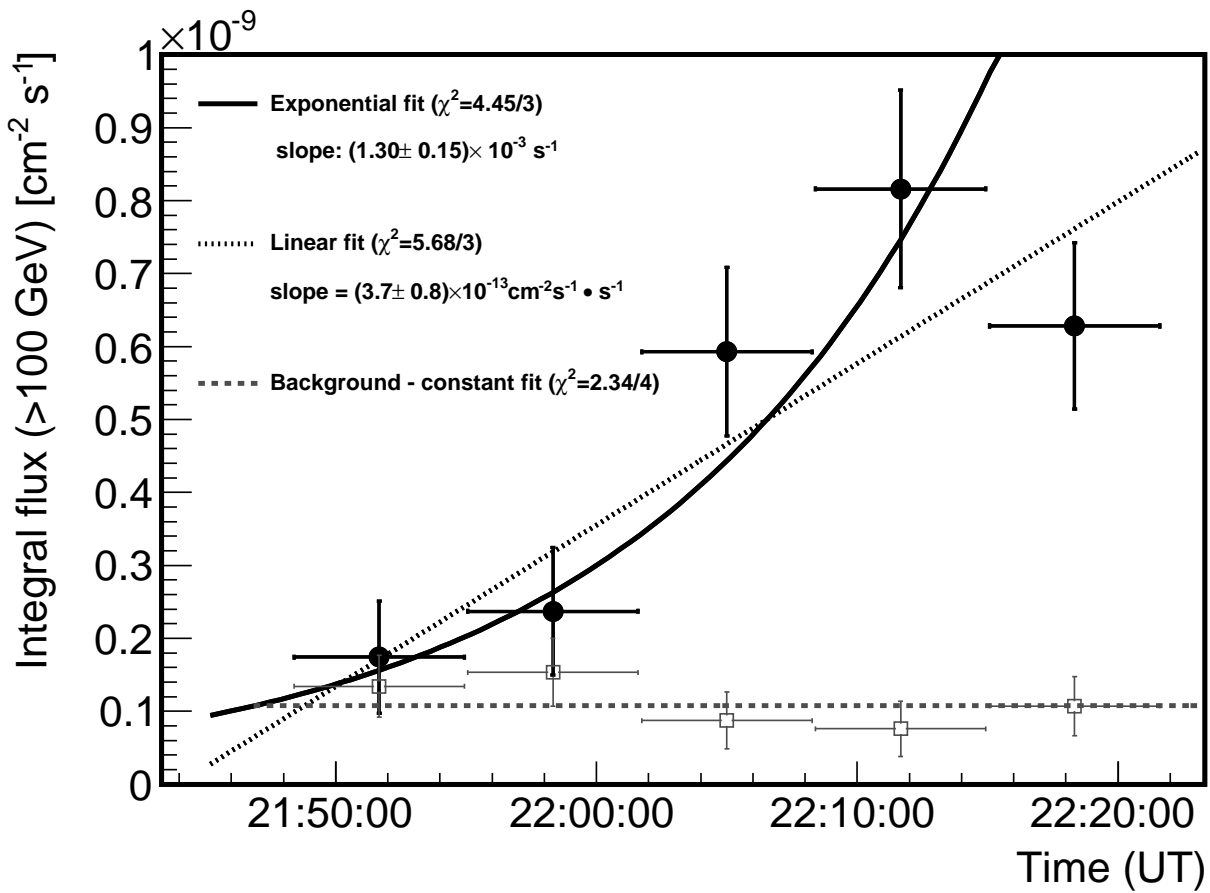}
\caption{PKS\,1222+21 light curve above 100\,GeV, in 6 minutes bins (black filled circles). The observation was carried out on MJD 55364.  The black solid line is a fit with an exponential function and the black dotted line a fit with a linear function. The grey open squares denote the fluxes from the background events and the grey dashed line is a fit with a constant function to these points.}
\label{fig:LC}
\end{figure}

\subsection{Light curve}

Despite the short observation time, the strength of the signal allows us to perform a variability study of the measured integral fluxes above 100\,GeV. 
The light curve binned in 6 minutes long intervals is shown in Fig.~\ref{fig:LC} and reveals clear flux variations. 
The constancy hypothesis  ($\chi ^2/{NDF} = 28.3/4$) is rejected  with high confidence (probability $<1.1\times 10^{-5}$). 
The fluxes of background events surviving the $\gamma$/hadron selection cuts are compatible with being constant and hence we can exclude a variation of the instrument performance during the observation.

To quantify the variability time scale we performed an exponential fit (solid black line  in Fig.~\ref{fig:LC}). A linear fit is also acceptable but does not allow us to define a time scale unambiguously.
For the exponential fit  the doubling time of the flare is estimated  as $8.6^{+1.1}_{-0.9}$\,minutes.
The derived timescale corresponds to the fastest time variation ever observed in a FSRQ in the VHE range and in any other energy range~\citep{Foschini2011}, and is among the shortest timescales measured on TeV emitting sources~\citep{HESS2155(2010)}.

\subsection{The HE -- VHE SED}

\begin{figure}[tbp]
\plotone{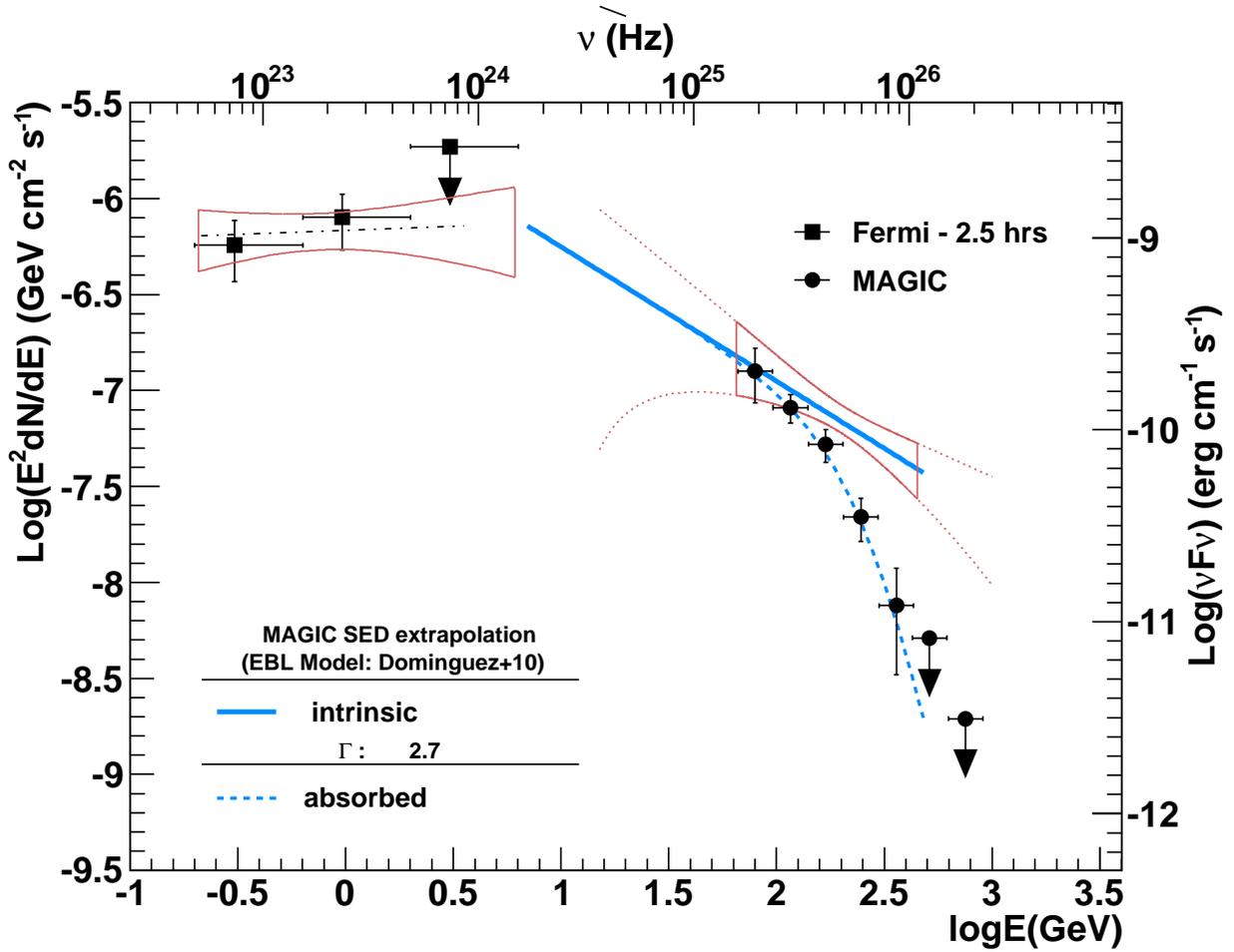}
\caption{High energy SED of PKS\,1222+21 during the flare of 2010 June 17 (MJD 55364.9), showing  \textit{Fermi}/LAT (squares) and MAGIC (circles) differential fluxes. 
A red bow tie in the MeV/GeV range represents the uncertainty of the likelihood fit to the \textit{Fermi}/LAT data.
The unfolded and deabsorbed spectral fit of the MAGIC data is also shown as a red bow tie, extrapolated to lower and higher energies (dotted lines) according to \citet{AbdoTeVAGN}. A thick solid line (photon index $\Gamma=2.7$) 
indicates a possible extrapolation of the MAGIC deabsorbed data to lower energies. 
The thick dashed line represents the EBL absorbed spectrum obtained from the extrapolated intrinsic spectrum using  the model  by \citet{Dominguez2010}. 
}
\label{fig:SED}
\end{figure}

In the high energy (HE) MeV/GeV energy range measured by \textit{Fermi}/LAT the source showed a significant flare lasting $\sim$3 days, with a flux peak on 2010 June 18 (MJD 55365)~\citep{Fermi draft}. 
 A dedicated analysis found that the 1/2\ h  MAGIC observation fell
within a gap in the LAT exposure, thus we analyzed a period of  2.5\,h (MJD 55364.867 to 55364.973), encompassing  the MAGIC observation.
 The LAT analysis for this time bin was performed as in~\citet{Fermi draft}, where
details can be found.
 It results in an integral flux   $(6.5\pm1.9)\times10^{-6}\,\mbox{cm}^{-2}\,\mbox{s}^{-1}$ at energy E\,$>\,$100\,MeV.  
The observation in such a short time does not provide any detection with
\textit{Fermi}/LAT at E$>$2\,GeV. Two \textit{Fermi}/LAT spectral points up to 2\ GeV together with an upper limit at the 95\% C.L. in the range $2-6.3$\,GeV are combined
with the MAGIC data in the Spectral Energy Distribution (SED) shown in  Fig.~\ref{fig:SED}.  

The figure also shows bow ties representing uncertainties associated with the spectral fits. 
The \textit{Fermi}/LAT spectrum is best described by a single power-law with index of $1.95\pm0.21$. 
In the case of MAGIC data the bow tie refers to the ``intrinsic" source spectrum, i.e.\ to the observed spectrum 
corrected for EBL absorption, described in sec.\ 3.1.  An extrapolation of the intrinsic spectrum in the MAGIC range to lower  energies is also shown indicating that:  i) there is a potentially smooth connection between the  \textit{Fermi}/LAT and MAGIC  extrapolated data in the 3 to 10 GeV region, ii) the photon index steepens from
1.9 in the  \textit{Fermi}/LAT range to 2.7 in the MAGIC range.
These results agree with the analysis of wider temporal intervals during this
flare and during the whole active period, in which the source spectrum is well described
by a broken power law with an energy break falling between 1 and 3 GeV~\citep{Fermi draft}.
Furthermore it is found that the high energy tail  (E\,$>2\,$GeV) of the \textit{Fermi}/LAT spectrum of PKS\,1222+21 extends up to 50 GeV, with a photon index in the range 2.4-2.8. 

\section{Discussion}

\subsection{EBL limits}

The interaction of very high energy $\gamma$-rays with low energy photons of the isotropic EBL  is a process with an energy dependent threshold, thus leading to an imprint of the EBL  density on the measured VHE $\gamma$-ray spectra of extragalactic sources~\citep{mazinraue2007}.
For PKS\,1222+21 ($z=0.432$), the measured spectrum spans from 70\,GeV to 400\,GeV probing EBL photons in the range 0.1 - 1\,$\mu$m (i.e.\ UV to near infrared). 

The EBL constraints using VHE $\gamma$-rays are usually derived assuming an intrinsic spectrum of the source \citep[e.g.][]{aharonian:2006:hess:ebl:nature}. 
In FSRQs, the presence of dense radiation fields of soft photons can lead to the internal absorption of VHE $\gamma$-rays, mimicking harder-than-intrinsic spectra  \citep[e.g.][]{SitarekBednarek2008}. However, for realistic spectral distributions of the internal photon fields it should not change the EBL limits significantly \citep{TavecchioMazin2009}. 

In our case the simultaneous data from \textit{Fermi}/LAT, which is free from internal or external absorptions, has been used to constrain the intrinsic photon index in VHE \citep[e.g.][]{Georganopoulos2010, FinkeRazzaque2009}.
We adopt a method similar to the one utilized by \citet{Georganopoulos2010}: the intrinsic spectrum in the VHE regime is assumed to follow the extrapolation of
the \textit{Fermi}/LAT above 3 GeV with a $\Gamma =2.4$. This is a conservative assumption since in reality the
spectrum could  soften with increasing energy.

\begin{figure}[tbp]
\plotone{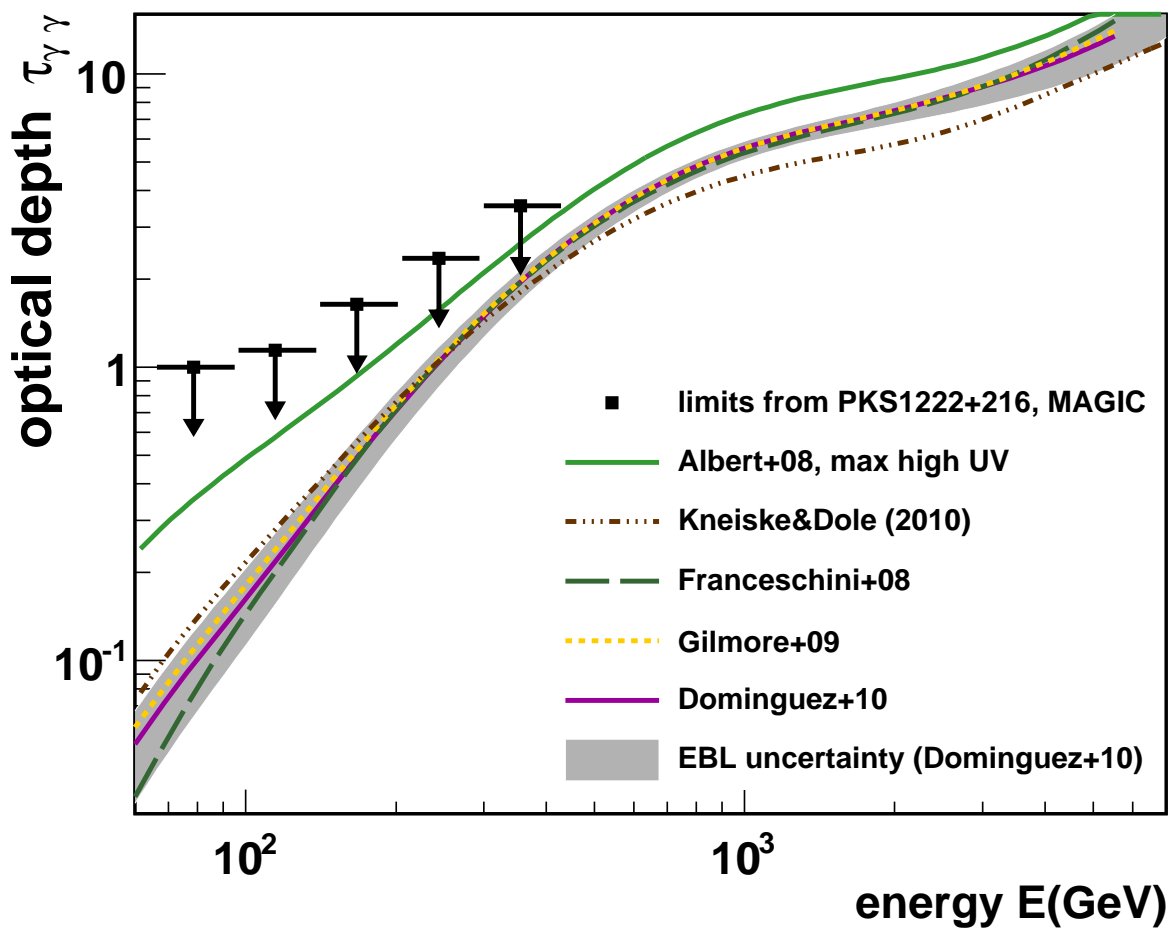}
\caption{Optical depth along the line of sight to PKS\,1222+21 (z=0.432) for a range of EBL models and the limits (95\% C.L.) from the MAGIC measurement,  assuming the limiting intrinsic photon index  $\Gamma_{\rm \tiny\mbox{VHE}}=2.4$. 
The grey-shaded area shows the uncertainties in the EBL determination as derived in \citet{Dominguez2010}, sec.~6.1 and Fig.~13.}
\label{fig:tau}
\end{figure}

The upper limit (95\% C.L.) on the
optical depth, $\tau_{\tiny\mbox{max}}$, for VHE $\gamma$-rays can be obtained from:
\begin{equation}
\tau_{\tiny \mbox{max}}(E) = \log \left[ \frac{F_{\tiny \mbox{intr}} (E)}{F_{\tiny \mbox{obs}}(E)-1.64\cdot \Delta F(E)}\right],
\end{equation} 
where $F_{\tiny \mbox{intr}} (E)$ is the maximum intrinsic flux at energy $E$,
$F_{\tiny \mbox{obs}}(E)$ and $\Delta F(E)$ are the MAGIC measured flux and its error, respectively.
The maximum intrinsic flux has been normalized at 70\,GeV assuming the EBL model giving a maximum flux absorption of ~30\% \citep{MAGIC3c279}.
The derived limits on the optical depth are shown in Fig.~\ref{fig:tau} 
together with a compilation of the predicted optical depths 
for a source at $z=0.432$ computed according to recent EBL models.  
The limits confirm previous constraints on the EBL models in the UV to near
infrared regimes derived using VHE
\citep{aharonian:2006:hess:ebl:nature,mazinraue2007,MAGIC3c279}
and HE spectra \citep{fermiEBL}.
Given the fact that the EBL models predict for
this redshift a stronger absorption with increasing energy, our data
do not indicate a softening of the spectrum within the energy range of our
observations.

\subsection{VHE $\gamma$-ray emission}

In the framework of the currently accepted EBL models, 
the observed {\em simultaneous} VHE and GeV spectra are consistent with a single power law with index 
$\sim 2.7\pm0.3$ between 3 GeV to 400 GeV, without a strong intrinsic cutoff. 
This evidence suggests that the 100 MeV - 400 GeV emission belongs to a unique component, peaking at
 $\approx 2-3\,$GeV, produced in a single region of the jet. If the emission process is inverse Compton scattering 
 on external photons by  relativistic electrons in the jet, as commonly assumed, a strong softening of the spectrum is expected above few tens of  GeV if the external photons derive from the BLR. This is due to the combination of two effects: the decreased  efficiency of the IC scattering occuring in  the Klein-Nishina (KN) regime \citep[e.g.][]{GhiselliniTavecchio2009} and the absorption of $\gamma$-rays through 
 pair production \citep{Reimer2007, TavecchioMazin2009, LiuBai2006}
\footnote{We note that this absorption has been invoked by \citet{Poutanen2010} to explain the existence of 
an apparently ``universal" break energy in the $\gamma$-ray spectrum of FSRQs at 2 GeV.}. 

The energy above which the KN effects become important can be roughly expressed as: $E_{\rm KN}\simeq 
22.5\,\nu_{o,15}^{-1}\,$GeV, where $\nu _{o,15}$ is the frequency of the target photons in units of $10^{15}$ 
Hz (or $E_{\rm KN} \simeq 75\,\lambda_{\mu {\rm m}}$ GeV in wavelength $\mu\rm{m}$ units). $\gamma$-ray 
absorption becomes effective when $E_{\rm \gamma \gamma} \simeq 60\,\nu _{o,15}^{-1}$ GeV 
($E_{\rm \gamma \gamma} \simeq 200\, \lambda_{\mu {\rm m}}$ GeV). Above this energy a cut-off is then expected.
The importance of both effects in the 10-100 GeV band is reduced if the external photon field is associated 
with the IR torus ($\nu_{\rm o}=10^{13}\,$Hz), as envisioned by the ``far dissipation" scenarios 
\citep[e.g.][]{Sikora2008}. In that case both effects start to be important above $\approx$ 1\,TeV. 
The absence of a spectral break or cutoff in the spectrum observed by MAGIC, strongly suggests that the $\gamma-$ray emission is not produced within the BLR.

The other important result of the MAGIC observation is the  evidence of  fast variability, 
$t_{\rm var}\sim10\,$minutes, indicating an extremely compact emission region with transverse dimensions, 
$R\sim 1.3\times10^{14}(\delta/10) (t_{\rm var}/10\,{\rm minutes})$ cm. This seems to be difficult to reconcile
with the ``far dissipation" scenarios if the emission takes place in the entire 
cross section of the jet \citep[see also][]{Tavecchio2010}. 
Estimating the size of the BLR, $R_{\rm BLR}$  from the accretion disk luminosity,  
$L_{\rm disk}=5\times 10^{45}$erg\,s$^{-1}$ \citep{Fan et al. 2006}, 
the distance of the emitting region is expected to be around $d>R_{\rm BLR}=3\times10^{17}$ cm. 
Assuming a conical jet with constant opening angle $\theta_{\rm j}$  \citep[see however, the suggestion of recollimation,][]{Marscher1980}, its size would be $R\sim\theta_{\rm j}d\sim3\times 10^{16} (\theta _{\rm j}/5{^\circ})$ cm. The absence of absorption features in the VHE spectrum allows also to exclude absorption within the emitting region and, together with the observed variability, to put a lower limit to the Doppler factor of the source. From \citet{Dondi1995}, Equation (3.7), assuming a power-law photon index 1.5 for the  spectrum of the optical target photons, we get a lower limit $\delta>15$, in agreement with Doppler factors derived from radio observations (Section 1).

A possibility to reconcile the spectral information (pointing to emission beyond the BLR) and the fast 
variability is to invoke the presence of very compact emission regions embedded within the large scale jet, as 
already proposed by several authors to explain the exceptionally rapid variability in PKS\,2155-304, Mkn\,501 
and AO\,0235+164 \citep{GhiselliniTavecchio2008, Giannios2009, Marscher2010}. An alternative possibility 
is that the jet experiences a strong recollimation forming a small emitting nozzle 
\citep[e.g.][]{NalewajkoSikora2009} as already suggested for M87 and PKS\,2155-304 
\citep[e.g.][]{BrombergLevinson2009, Stawarz2006}.
Alternative scenarios involving proton-driven cascades or proton-synchrotron emission in amplified magnetic fields, e.g. generated by filamentation instabilities~\citep{Jacob2010}, could also play a role.

In conclusion the MAGIC observations of VHE emission from the FSRQ PKS 1222+21 put severe constraints on 
emission models of blazar jets. These results were obtained from a short observation of a flaring source
thanks to the collaboration between the MAGIC and \textit{Fermi} projects. Repeated and hopefully longer observations
of flaring blazars with MAGIC and \textit{Fermi} promise substantial progress in the study of extreme blazars.
 
\acknowledgments
\section*{Acknowledgments}
We thank the Instituto de Astrof\'{\i}sica de
Canarias for the excellent working conditions at the
Observatorio del Roque de los Muchachos in La Palma.
The support of the German BMBF and MPG, the Italian INFN, 
the Swiss National Fund SNF, and the Spanish MICINN is 
gratefully acknowledged. This work was also supported by 
the Marie Curie program, by the CPAN CSD2007-00042 and MultiDark
CSD2009-00064 projects of the Spanish Consolider-Ingenio 2010
programme, by grant DO02-353 of the Bulgarian NSF, by grant 127740 of 
the Academy of Finland, by the YIP of the Helmholtz Gemeinschaft, 
by the DFG Cluster of Excellence ``Origin and Structure of the 
Universe", and by the Polish MNiSzW Grant N N203 390834.

The $Fermi$/LAT Collaboration acknowledges support from a number of
agencies and
institutes for both development and the operation of the LAT as well
as scientific
data analysis. These include NASA and DOE in the United States, CEA/Irfu and
IN2P3/CNRS in France, ASI and INFN in Italy, MEXT, KEK, and JAXA in
Japan, and the
K.~A.~Wallenberg Foundation, the Swedish Research Council and the
National Space
Board in Sweden. Additional support from INAF in Italy and CNES in France for
science analysis during the operations phase is also gratefully acknowledged.

\clearpage

\end{document}